\documentstyle[preprint,aps,epsbox]{revtex}

\newcommand{\mpl}{m_{\rm Pl}}
\newcommand{\etal}{{\it et al.}}

\begin{document}

\draft
\preprint{YITP-99-52, KUNS-1597, gr-qc/9909027}
\title{Causal structure of an inflating magnetic monopole}
\author{Nobuyuki Sakai\thanks{Electronic address: sakai@yukawa.kyoto-u.ac.jp}}
\address{Yukawa Institute for Theoretical Physics, Kyoto University, Kyoto
606-8502, Japan}
\author{Ken-ichi Nakao}
\address{Department of Physics, Osaka City University, Osaka 558-8585, Japan}
\author{Tomohiro Harada}
\address{Department of Physics, Kyoto University, Kyoto 606-8502, Japan}
\date{1 December 1999}
\maketitle

\begin{abstract}

We clarify the causal structure of an inflating magnetic monopole.
The spacetime diagram shows explicitly that this model is free from ``graceful
exit'' problem, while the monopole itself undergoes ``eternal inflation''.
We also discuss general nature of inflationary spacetimes.

\end{abstract}

\vskip 1cm
\begin{center}
PACS number(s): 04.70.Bw, 14.80.Hv, 98.80.Cq
\end{center}

\newpage
\baselineskip = 19pt

For the last decade spacetime solutions of magnetic monopoles have
been intensively studied in the literature \cite{VG,OLB,TMT,LV,Sak,BTV}.
This originated from the rather mathematical interest in static
solutions with non-Abelian hair \cite{VG}. It was shown \cite{OLB} that
static regular solutions are nonexistent if the vacuum expectation
value of the Higgs field $\eta$ is larger than a critical value
$\eta_{\rm sta}$, which is of the order of the Planck mass $\mpl$.

Linde and Vilenkin independently pointed out that such
monopoles could expand exponentially in the context of inflationary
cosmology \cite{LV}.
Because this ``topological inflation'' model does not require
fine-tuning of the initial conditions, it has been attracting
attention. In particular, it is recently found that topological inflation
takes
place in some of the plausible models in particle physics \cite{EIS}.

In Ref.\cite{Sak} (Paper I), dynamical solutions of magnetic monopoles
for $\eta>\eta_{\rm sta}$ were numerically obtained:
monopoles actually inflate if $\eta\gg\eta_{\rm sta}$.
Recently, the causal structure of an inflating magnetic monopole was
discussed in Ref.\cite{BTV}.
The spacetime diagrams in Refs.\cite{BTV,VT} showed, for instance, that
the inflationary boundary expands along outgoing null geodesics,
that is, any observer cannot exit from an inflationary region.

The above argument would be fatal to topological inflation because it
implies that reheating never occurs. Therefore, the spacetime
structure of topological inflation deserves close examination.
In this paper we clarify the causal structure of an inflating
magnetic monopole, as a complement of Paper I.
As a result, we find that the diagrams in Refs.\cite{BTV,VT} are
incorrect, as we shall show below.

In order to see the spacetime structure for numerical solutions, we observe
the signs
of the expansion of a null geodesic congruence. For a spherically
symmetric metric,
\begin{equation}\label{metric}
ds^2=-dt^2+A^2(t,r)dr^2+B^2(t,r)r^2(d\theta^2+\sin^2\theta d\varphi^2),
\end{equation}
an outgoing $(+)$ or ingoing $(-)$ null vector is given by
\footnote{
\baselineskip = 18pt
In Paper I and Ref.\cite{SSTM}, the null vector was expressed as
$k^{\mu}_{\pm}=(-1,\pm A^{-1},0,0)$. The minus sign of $k_{\pm}^t$ was
just a typo, and the expression of $\Theta_{\pm}$, which is the same
as Eq.(\ref{expansion}), was correct.}
\begin{equation}\label{null}
k^{\mu}_{\pm}=(1,\pm A^{-1},0,0)
\end{equation}
and its expansion $\Theta_{\pm}$ is written as
\begin{equation}\label{expansion}
\Theta_{\pm} =
{k^{\theta}_{\pm;\theta}}+{k^{\varphi}_{\pm;\varphi}}
={2\over B}\bigg({\partial B\over\partial t}
\pm{1\over Ar}{\partial (Br)\over\partial r}\biggl),
\end{equation}
which is defined as the trace of a {\it projection} of $k^{\mu}_{;\nu}$
onto a relevant 2-dimensional surface. The derivation of Eq.(\ref{expansion})
as well as more general arguments were given by Nakamura \etal\cite{NOK}~
We define an ``apparent horizon'' as the surface with $\Theta^+=0$ or
$\Theta^-=0$
\footnote{
\baselineskip = 18pt
In the literature an ``apparent horizon'' usually refers to the outermost
surface
with $\Theta^{+}=0$ in an asymptotically flat spacetime.
In this article, however, we call any marginal surface with
$\Theta^+=0$ or $\Theta^-=0$ an apparent horizon.}.
We label those surfaces as S1, S2, \ etc. in our figures.

In Fig.\ 1(a), we plot the trajectories of monopole boundaries and of apparent
horizons in terms of the proper distance from the center: $X\equiv\int^r_0
Adr'$.
Here we define the boundary in two ways: $X_{\Phi}$ as the
position of $\Phi=\eta/2$ and $X_w$ as the position of $w=1/2$, where
$\Phi$ and $w$ are the Higgs field and a gauge-field function, respectively.
An apparent horizon S1 almost agrees with $X=H_0^{-1}\equiv[8\pi
V(0)/3\mpl^2]^{-1/2}$,
which implies that the monopole core is almost de Sitter spacetime.
Figure 1(a) also illustrates that the monopole actually inflates.

Based on the solution in Fig.\ 1(a), we schematically depict the
embedding diagram in Fig.\ 2.
We see that wormhole structure with black hole horizons appears
around an inflating core. Because the inflating core becomes causally
disconnected from the outer universe, such an isolated region is
called a ``child universe''. The production of child universes were
originally discussed by Sato, Sasaki, Kodama, and Maeda (SSKM) \cite{SSKM}
in the context of the original inflationary model associated with a
first-order phase
transition \cite{SG}.
In monopole inflation, similar spacetime structure appears with a
simpler potential (as in Fig.\ 5) and with natural initial conditions.

In order to see the causal structure of the inflating monopole more
closely, we also plot ingoing and outgoing null geodesics in Fig.\ 1(b).
The similar diagram for an inflating global monopole was presented by Cho and
Vilenkin \cite{CV}.
An important feature for the magnetic monopole as well as the global monopole
is that the boundary moves {\it inward} in terms of the coordinate $r$
and, moreover, it eventually becomes {\it spacelike}.

To understand the global structure, which is not completely covered by the
numerical solution, we make a reasonable assumption that the
core region approaches to de Sitter spacetime and the outside to
Reissner-Nordst\"{o}rom, as in Refs.\cite{TMT,BTV}.
In each static spacetime, apparent horizons and event horizons are
identical; the signs of $(\Theta^+,\Theta^-)$ are determined
as is shown in Fig.\ 3.
The above assumption implies that, although the whole space in the
early stage is quite dynamical and an apparent horizon does
not coincide with an event horizon, the two horizons later approach
to each other. Hence, we can extrapolate the
global structure from the structure of apparent horizons in the local
numerical solution.
From the consistency of the spatial distribution of the signs of
$(\Theta^+,\Theta^-)$, we conclude that Fig.\ 3 gives the only possible
embedding.

It is also instructive to compare the inflating monopole solution with
SSKM's model of child universe. For SSKM model, Blau \etal\ studied the
details of
the boundary motion and the causal structure under the thin-wall approximation
\cite{BGG}. Figure 3 is quite similar to that for their Type E
solution, except that the inflationary boundary in SSKM model is lightlike
or timelike. The similarity confirms our results in Figs.\ 1-3, and the
difference is also easily understood: the boundary in SSKM model
is a domain wall, or a solitonic wave, and hence it cannot be spacelike.

On the other hand, Fig.\ 4 of Borde, Trodden, and Vachaspati \cite{BTV}
or Fig.\ 3 of Vachaspati and Trodden \cite{VT} are quite different
from our fig.\ 3 in the following respects.
\begin{itemize}
\item The inflationary boundary is expressed as {\it outgoing null
geodesics} (as in Fig.\ 5(a)).
\item The horizon structure is inconsistent with the monopole solution
in Fig.\ 1.
\item There is no de Sitter region in the initial time hypersurface.
\end{itemize}
Their diagrams were not based on any analytic nor numerical
solution, and there is no ground that such a spacetime satisfies
Einstein equations.
Instead they claimed that the horizon argument in Paper I was incorrect
for the following reason \cite{BTV}.
$k^{\mu}_{\pm}$ {\it in Eq.}(\ref{null}) {\it is not the tangent vector
associated with affinely parameterized radial null geodesics, and so
its expansion $\Theta\equiv k^{\mu}_{:\mu}$ cannot be used to find trapped and
anti-trapped surfaces.}

The above statement is, however, untrue for the following reason. Their
misunderstanding stemmed from different definitions of the
expansion $\Theta$. If we adopted the definition $\Theta_{\rm BTV}\equiv
k^{\mu}_{;\mu}=k^{t}_{;t}+k^{r}_{;r}+k^{\theta}_{;\theta}+k^{\varphi}_{;\
varphi}$,
the null vector $k^{\mu}_{\pm}$ should be chosen so as to satisfy the
geodesic equation.
However, it is different from our expression (\ref{expansion}), $\Theta
=k^{\theta}_{;\theta}+k^{\varphi}_{;\varphi}$.
With this expression, it is irrelevant whether $k^{\mu}_{\pm}$ is a solution
of
null geodesic equations or not, as we explain as follows.
The outgoing $(+)$ or ingoing $(-)$ null vector associated radial null
geodesics is generally expressed as
$k^{\mu}_{\pm}=f(1,\pm A^{-1},0,0)$, where $f$ is the positive
function which is determined by the null geodesic equations.
One can easily check $\Theta_{\pm}$ is simply changed into $f\Theta_{\pm}$ as
$k^{\mu}_{\pm}\rightarrow fk^{\mu}_{\pm}$. That is, the sign of
$\Theta$ can be determined without solving null geodesic equations,
as long as one chooses the signs for vector components appropriately.
Thus their criticism for Paper I is falsified.

Cosmologically, Fig.\ 3 tells us that any inflationary region eventually
enters a reheating phase because any observer inside the core
finally goes out. On the other hand, the monopole boundary continues
to expand in terms of the physical size and approaches spatial
infinity ($i^0$).
Figure 3 thus proves that this model free from ``graceful exit'' problem,
while the monopole itself undergoes ``eternal inflation''.

Finally, we extend discussions to general nature of inflationary
spacetimes. What we have shown is that the boundary of an inflating
monopole always goes inward and is spacelike. We should notice here that this
behavior is not a special feature of monopole inflation but a general nature
of any slow-roll inflation. In any model of slow-roll inflation with 
a reheating phase, inflation ceases when the inflaton field $\Phi$ becomes 
larger (or smaller) than a critical value $\Phi_{\rm cr}$.
Therefore, the boundary of an inflationary region is characterized as a hypersurface 
with $\Phi=\Phi_{\rm cr}=\mbox{const}$. Whether a $\Phi=\mbox{const}$ 
hypersurface is spacelike, timelike or null
is up to the sign of $\nabla_{\mu}\Phi\nabla^{\mu}\Phi$. 
In general spacetimes, the line element is given as
\begin{equation}
ds^2=-\alpha^2 dt^2+\gamma_{ij}(dx^{i}+\beta^{i}dt)(dx^{j}+\beta^{j}dt),
\end{equation}
where $\alpha$ and $\beta^{i}$ are the lapse function and the shift vector,
respectively.
Then, $\nabla_{\mu}\Phi\nabla^{\mu}\Phi$ is written as
\begin{equation}\label{normal}
\nabla_{\mu}\Phi\nabla^{\mu}\Phi=-\frac{1}{\alpha^{2}}
\left(\frac{\partial\Phi}
{\partial t}\right)^2+
2\frac{\beta^{i}}{\alpha^2}\left(\frac{\partial\Phi}
{\partial t}\right)\left(\frac{\partial\Phi}
{\partial x^{i}}\right)
+\left(\gamma^{ij}-\frac{\beta^{i}\beta^{j}}{\alpha^2}\right)
\left(\frac{\partial\Phi}
{\partial x^{i}}\right)\left(\frac{\partial\Phi}
{\partial x^{j}}\right).
\end{equation}
According to the standard picture of inflation, 
the scalar field becomes almost homogeneous
in a sufficiently inflated region.
It follows that there exists a time slicing such that
the second and the third terms in the right-hand-side of (\ref{normal}) are
negligibly small compared with the first term.
Therefore, $\nabla_{\mu}\Phi\nabla^{\mu}\Phi$ is negative
in the sufficiently inflated region, which implies that the 
$\Phi=\mbox{const}$ hypersurface must be {\it spacelike}.

Now that we have shown the boundary of an inflating region is 
spacelike, it is obvious that it moves not {\it outward} but {\it inward}.
Consider, for example, the Higgs potential $V(\Phi)$ as shown in Fig.\ 
4. If the boundary moved outward, any 
observer would see that the field climbs the potential up from $\Phi\cong\eta$ 
to $\Phi\cong0$ and never rolls down to $\Phi\cong\eta$; however, such 
classical dynamics is improbable.
We therefore conclude that the causal structure of an inflating local
region is schematically described by Fig.\ 5(b). Although one may intuitively 
draw a diagram like Fig.\ 5(a), or
Fig.\ 2 of Vachaspati and Trodden \cite{VT}, it is incorrect.

\acknowledgements
The numerical computations of this work were carried out at the
Yukawa Institute Computer Facility.
N.S.\ and T.H.\ were supported by JSPS Research Fellowships for Young
Scientists, No.\ 9702603 and No.\ 9809204.


\end{document}